\documentclass[aps,prd,floats,floatfix,amssymb,twocolumn,superscriptaddress]{revtex4-1}

\usepackage{amsmath}
\usepackage{amsfonts,amssymb}
\usepackage{graphicx,graphics}
\usepackage{hyperref}
\usepackage{ulem}
\usepackage{color}

\DeclareMathOperator\arctanh{arctanh}

\begin{document}

\title{Unruh-DeWitt detector in $\textrm{AdS}_2$}

\author{Jo\~ao P. M. Pitelli}
\email[]{pitelli@unicamp.br}
\affiliation{Departamento de Matem\'atica Aplicada, Universidade Estadual de Campinas,
13083-859 Campinas, S\~ao Paulo, Brazil}%

\author{Bruno S. Felipe}
\email[]{brunfeli@ifi.unicamp.br}
\affiliation{Instituto de F\'isica ``Gleb Wataghin'', Universidade Estadual de Campinas, 13083-859 Campinas, S\~ao Paulo, Brazil}

\author{Ricardo A. Mosna}
\email[]{mosna@unicamp.br}
\affiliation{Departamento de Matem\'atica Aplicada, Universidade Estadual de Campinas,
13083-859 Campinas, S\~ao Paulo, Brazil}%

\begin{abstract}
We find the response function and the transition rate for an Unruh-DeWitt detector interacting with a conformal  scalar field in global two-dimensional anti--de Sitter (AdS) spacetime with different boundary conditions at its conformal infinities. We calculate the particle energy spectrum as seen by subcritical accelerated detectors and discuss how it depends on the choice of the  boundary condition. We show that, despite this nontrivial dependence on the boundary conditions, the limit when the AdS length scale tends to zero is well defined and leads to the well-known results of 1+1 Minkowski space. One can thus interpret the AdS energy scale as a natural regulator for the well-known infrared ambiguity of massless scalar fields in 1+1 Minkowski spacetime.
\end{abstract} 

\maketitle

\section{\label{sec:intro}Introduction}

The theory of quantum fields in curved spacetime provides a well-established prescription for the quantization of fields propagating in a classical background provided by general relativity. In this context, the spacetime (a solution of the Einstein equations) is treated classically, while the field propagating in this background is quantized neglecting the possible effects of backreaction.

In this setup, the concept of particles does not have a universal meaning, being an observer-dependent concept~\cite{birrell}. In particular, as stated by Unruh in Ref.~\cite{unruh}, the observation of particles by a particle detector depends on its state of motion.  Since the Minkowski spacetime is maximally symmetric, we usually choose the Poincar\'e invariant state, $\left|0\rangle\right.$, defined by 
\begin{equation}
a_{\vec{k}}\left|0\rangle\right.=0,\,\,\, \forall\, \vec{k},
\end{equation}
as a ``natural'' vacuum. Here, the field $\phi(x)$ is expanded in terms of Minkowski plane wave modes $u_{\vec{k}}(x)$ as 
\begin{equation}
\phi(x)=\sum_{\vec{k}}{\left[a_{\vec{k}}u_{\vec{k}}(x)+a_{\vec{k}}^{\dagger}u_{\vec{k}}^{\ast}(x)\right]}.
\end{equation}
It follows that this $\left|0\rangle\right.$ describes the same  vacuum state for every inertial observer.

Anti--de Sitter (AdS) spacetime is  maximally symmetric as well, having the highest possible degree of symmetries. Inspired by the above construction, perhaps one could assume to be straightforward to define an AdS invariant vacuum state $\left|\Psi\rangle\right.$ that would be common to all  inertial observers. However, there is a crucial difference between Minkowski and AdS. AdS is not globally hyperbolic and  the evolution of fields in this background is not uniquely defined given the initial data on any of its spacelike surfaces. In fact, AdS has a conformal timelike boundary, where boundary conditions must be imposed on the fields in order for them to have a well-defined evolution.

In this paper, we adopt Wald's prescription  for the evolution of fields in nonglobally hyperbolic spacetimes~\cite{wald,waldishibashi,waldishibashi2}. In this setup, the possible sensible dynamics for the Klein-Gordon field are in one-to-one correspondence with the positive self-adjoint extensions of the spatial part of the wave operator. Parametrizing the self-adjoint extensions by $\beta$, the field can be expanded in terms of the normal modes $u_{\vec{k}}^{\beta}$ satisfying the boundary condition associated with $\beta$ as
\begin{equation}
\phi^{\beta}(x)=\sum_{\vec{k}}{\left[a^{\beta}_{\vec{k}}u^{\beta}_{\vec{k}}(x)+a_{\vec{k}}^{\beta\dagger}u_{\vec{k}}^{\beta\ast}(x)\right]}.
\end{equation}
Once we define the vacuum state $\left|0\rangle_{\beta}\right.$ by
\begin{equation}
a_{\vec{k}}^{\beta}\left|0\rangle_\beta\right.=0,\,\,\, \forall\, \vec{k},
\end{equation}
it will depend explicitly on $\beta$ and may even be no longer AdS invariant~\cite{pitelli1,pitelli2}.

We work for simplicity on global $\textrm{AdS}_2$, whose metric is given by
\begin{equation}
ds^2=\frac{L^2}{\cos^2{\rho}}(-dt^2+d\rho^2).
\end{equation}
We investigate the response function and the transition rate of a detector following trajectories with constant $\rho$, which correspond to inertial ($\rho=0$) and uniformly accelerated ($\rho\neq 0$) motion in AdS$_2$. 
As shown in Refs.~\cite{deser,jennings}, there is a threshold for this acceleration above which the temperature measured by the detector is well defined. In what follows, we study accelerated observers below this threshold, so that no thermal spectrum is seen by the particle detector.  After studying the detector's response in AdS$_2$, we carefully analyze its $L\to\infty$ limit and relate it to the response of inertial detectors in Minkowski space. 

It is well known that a massless scalar field in Minkowski spacetime is infrared (IR) divergent. This divergence can be controlled by an IR cutoff $m_0$~\cite{mink1+1}, which introduces an ambiguity in the Wightman function and hence in the detector's response function. Nevertheless, it can be shown that it is possible to extract a physical result (independent of $m_0$) for the transition rate by considering {\it ad hoc} regularization methods (see, e.g., Ref.~\cite{tese}). One of the aims of this work is to show that AdS$_2$ works as a natural regulator for 1+1 Minkowski spacetime.  In fact, we show that the inverse length $1/L$  is closely related to $m_0$ irrespective of the choice of boundary conditions in AdS. It is interesting to note that by considering the 1+1 Minkowski spacetime under this kind of limiting procedure, the  
arbitrary UV regulator $\epsilon>0$ is also irrelevant. In this way, both the IR and UV divergences are dealt with at once by this procedure, which may be summarized by ``get the transition rate for AdS$_2$ and take the $L\to\infty$ limit''.

This paper is organized as follows: In Sec.~\ref{sec:minkowski}, we briefly review a few concepts about Unruh-DeWitt detectors in curved spacetimes. In particular, we  rederive the response function for the  massless scalar field  in 1+1 Minkowski spacetime for the case when the detector is abruptly switched on and off. In Sec.~\ref{sec:bc}, we motivate our choices for the boundary conditions at the  conformal boundaries in AdS$_2$. In Sec.~\ref{sec:wightman}, we derive expressions for the field modes and Wightman function for each one of these  choices. Sections~\ref{sec:response} and~\ref{sec:limits} present the main results of the paper, wherein we calculate the response function and the transition rates in AdS$_2$ and consider their $L\to\infty$ limit. In Sec.~\ref{sec:regularization}, we unveil the relation between $1/L$ and $m_0$ and discuss how this throws light on the results of the previous sections. Finally, in Sec.~\ref{sec:conlusion} we present our main conclusions.

\section{\label{sec:minkowski}Response function and transition rate in  1+1 Minkowski spacetime }

It is well known that the response function for the Unruh-DeWitt detector in a general curved spacetime  is given by~\cite{birrell}
\begin{equation}\begin{aligned}
    \mathcal{F}(\Omega) =\lim_{\epsilon\to 0^{+}} &\int_{-\infty}^{\infty}\mathrm{d}\tau\int_{-\infty}^{\infty}\mathrm{d}\tau'e^{-i\Omega(\tau-\tau')}\times\\&\times \chi(\tau)\chi(\tau^{\prime})W_\epsilon(x(\tau),x(\tau^{\prime})),
    \label{response function}
\end{aligned}\end{equation}
where $\Omega=E_f-E_i$ is the energy gap between the initial and final states of the detector, $x(\tau)$ is the detector's trajectory as a function of proper time, $\chi(\tau)$ is the switching function (which effectively turns the detector on and off), and $W_{\epsilon}(x,x^{\prime})=\langle\left. 0\right|\phi(x)\phi(x^{\prime})\left|0\right.\rangle$ is the Green-Wightman function with the standard regularization $t\to t-i \epsilon$. The response function given by Eq.~(\ref{response function})  is essentially determined by the Wightman function along the detector's trajectory and the switching function. When the Wightman function is invariant under time translation, i.e.,  when $W_\epsilon(x(\tau),x(\tau^{\prime}))=W_\epsilon(\Delta \tau)$  (the case of interest in this paper), it is useful to make a change of coordinates from ($\tau,\tau^\prime$) to ($u,s$), with  $u:=\tau$, $s:=\tau-\tau'$ when $\tau'<\tau$, and $u:=\tau'$, $s:=\tau'-\tau$ when $\tau<\tau'$. This leads to
\begin{equation}\begin{aligned}
    \mathcal{F}(\Omega) = 2\lim_{\epsilon\to 0^{+}}\int_{-\infty}^{\infty}\mathrm{d}u & \int_{0}^{\infty}\mathrm{d}s \,\chi(u)\chi(u-s)\times\\&\times Re\left[e^{-i \Omega s}W_\epsilon(s)\right].
\end{aligned}\end{equation}

An abrupt switching function given by $\chi(\tau)=\Theta(\tau-T_0)\Theta(T-\tau)$ represents a detector that is turned on at time $T_0$ and read at time $T$. The response function in this case is given by
\begin{equation}\label{resp}
    \mathcal{F}_{T}(\Omega) = 2\lim_{\epsilon\to 0^{+}}\int_{T_0}^{T}\mathrm{d}u\int_{0}^{\,u-T_0}\mathrm{d}s \,Re\left[e^{-i\Omega s}W_\epsilon(s)\right].
\end{equation}
The instantaneous transition rate $\dot{\mathcal{F}}_{T}(\Omega)$ is defined as the derivative of $\mathcal{F}_T(\Omega)$ with respect to $T$ and represents the number of clicks of the detector per unit time. It can be written as
\begin{equation}\label{rate}
   \dot{ \mathcal{F}}_{T}(\Omega) = 2\lim_{\epsilon\to 0^{+}}\int_{0}^{T-T_0}\mathrm{d}s \,Re\left[e^{-i\Omega s}W_\epsilon(s)\right].
\end{equation}
The choice of the switching function has of course a nontrivial effect on $\mathcal{F}(\Omega)$. In a $d$-dimensional spacetime with $d\geq 4$, an abrupt switching is known to lead to a divergent response function~\cite{matsas,svaiter,Hodgkinson}. However, for $3\leq d <6$, one can extract physically meaningful transition rates by (first) carefully removing  the UV $\epsilon>0$ regulator while maintaining a continuous switching function and (then) taking the sharp switching limit~\cite{Hodgkinson}. For $d=2$ the logarithmic behavior of the Wightman function makes the integrals (\ref{resp}) and (\ref{rate}) converge so that the UV regularization may be forgone (more on this below). There is an important caveat for $d=2$, though. The massless field is IR divergent and some other type of regularization procedure may be necessary, as we discuss next. Finally, to overcome possible transient effects due to the switching on process, it might be interesting to take $T_0\to-\infty$.


We now review some standard results for a  massless scalar field in 1+1  Minkowski spacetime. In this case, the Wightman function for a detector on an inertial path reads~\cite{mink1+1}
\begin{equation}\label{greenmink1}
        W_\epsilon^{\text{Mink}}(\Delta t)=-\frac{1}{2\pi}\ln{\left[m_0(\epsilon+i\Delta t)\right]},
\end{equation}
where $\epsilon>0$ is an UV regularization parameter and $m_0>0$ is  an IR frequency cutoff, which is required for massless fields. It is easy to show that
\begin{equation}
\left|e^{-i \Omega s }\ln{\left[m_0(\epsilon+i\Delta t)\right]}\right|\leq \sqrt{\left[\ln{(m_0 s)}\right]^2+\left(\frac{\pi}{2}\right)^2},
\end{equation}
and that the right-hand side of this equation is integrable. Hence, the limit $\epsilon\to 0^{+}$ can be taken under the integral sign by  dominated convergence and, by substituting Eq.~(\ref{greenmink1}) into Eq.~(\ref{resp}), we have
\begin{equation}\label{minkfinito}
    \begin{aligned}
       \mathcal{F}_T^{\text{Mink}}(\Omega) =-&\int_{T_0}^{T}du\int_{0}^{u-T_0}ds\bigg\{\frac{1}{2}\sin{(\Omega s)}\\&+\frac{1}{\pi}\cos{(\Omega s)}\ln{(m_0 s)}\bigg\}.
    \end{aligned}
\end{equation}
Integrating the above expression yields
\begin{equation}
 \begin{aligned}
       \mathcal{F}_T^{\text{Mink}}(\Omega)
        &=\frac{1}{2 \pi  \Omega^2}\Bigg\{2\gamma-2 +2\Delta T \Omega\, \text{Si}(\Delta T\Omega)-\\&2\text{Ci}(\Delta T\, \Omega) -\pi\Delta T\Omega+\pi \sin(\Delta T\Omega) +\\& 2\ln{(m_{0}^{-1}\Omega)}+2\cos(\Delta T\Omega)\left[1+\ln{(m_{0}\Delta T)}\right]\Bigg\},
       \end{aligned}
   \label{transitionfinal}
\end{equation}
with $\Delta T\equiv T-T_0$, $\textrm{Ci}(-|x|)\equiv i \pi+\textrm{Ci}(|x|)$, and $\ln(-|x|)\equiv i\pi+\ln{|x|}$. $\textrm{Si}(x)$ and $\textrm{Ci}(x)$ are the sine integral and cosine integral functions, respectively. Equation~(\ref{transitionfinal}) has an ambiguity given by the infrared regulator $m_0$. However, the average rate of transition $\mathcal{R}_T^{\text{Mink}}(\Omega)\equiv \frac{  \mathcal{F}_T^{\text{Mink}}(\Omega)}{\Delta T}$ is meaningful in the limit $\Delta T\to \infty$ and is given by
\begin{equation}
\mathcal{R}^{\text{Mink}}(\Omega)\equiv\lim_{\Delta T\to\infty}{\frac{  \mathcal{F}_T^{\text{Mink}}(\Omega)}{\Delta T}}=-\frac{1}{\Omega}\Theta(-\Omega).
    \label{ratefinal}
\end{equation}
One can also obtain the same result for $\dot{\mathcal{F}}_T(\Omega)$ using  Eq.~(\ref{rate}) in the limit $T_0\to-\infty$ by inserting an exponential cutoff $e^{-s/\delta}$ in place of the sharp switching limit~\cite{tese}. If one takes the limit $m_0\to 0$ first and then $\delta\to\infty$, the ambiguity due to $m_0$ is eliminated and Eq.~(\ref{ratefinal}) is recovered.

\section{\label{sec:bc}Boundary conditions and conformal fields in $\textrm{AdS}_2$}

As discussed in the Introduction, AdS is a nonglobally hyperbolic spacetime: when solving the wave equation, its solutions are not fully determined by the initial data. The evolution of a classical wave in $\textrm{AdS}$ depends crucially on the exchange of information with the conformal boundary, which can be modeled by an appropriate boundary condition for the field.

Here we follow Wald's approach~\cite{wald,waldishibashi,waldishibashi2} to tackle this problem. In this setup, the possible dynamics of a classical field are in one-to-one correspondence with the positive self-adjoint extensions of the spatial part of the wave operator. Although this is not the only possible prescription, it provides a very reasonable  dynamics respecting causality, time translation/time reflection invariance, and (what is most important) a conserved energy functional~\cite{waldishibashi}. Moreover, the positivity of the self-adjoint extensions imply stability, with no  generic solutions growing unboundedly in time, and the quantization process is straightforward.

The two-dimensional $\textrm{AdS}$ spacetime differs from its higher-dimensional counterparts in that  it possesses two disconnected boundaries. A conformal scalar field propagating in global $\textrm{AdS}_2$ thus behaves like a free field in a box. It can be shown that, in this case,  there are an infinite number of self-adjoint extensions to this problem parametrized by $U(2)$~\cite{bonneau}. In what follows, we will consider two classes of boundary conditions representing positive self-adjoint extensions: (i)  the first class is given by the most commonly used Robin boundary conditions; (ii) in the second class, we (effectively) close the spatial sections by imposing that the wave function at the two boundaries differs only by a phase. These two classes illustrate our main points with relatively simple calculations.

\begin{figure}[ht]
\centering
\includegraphics[width=0.9\columnwidth]{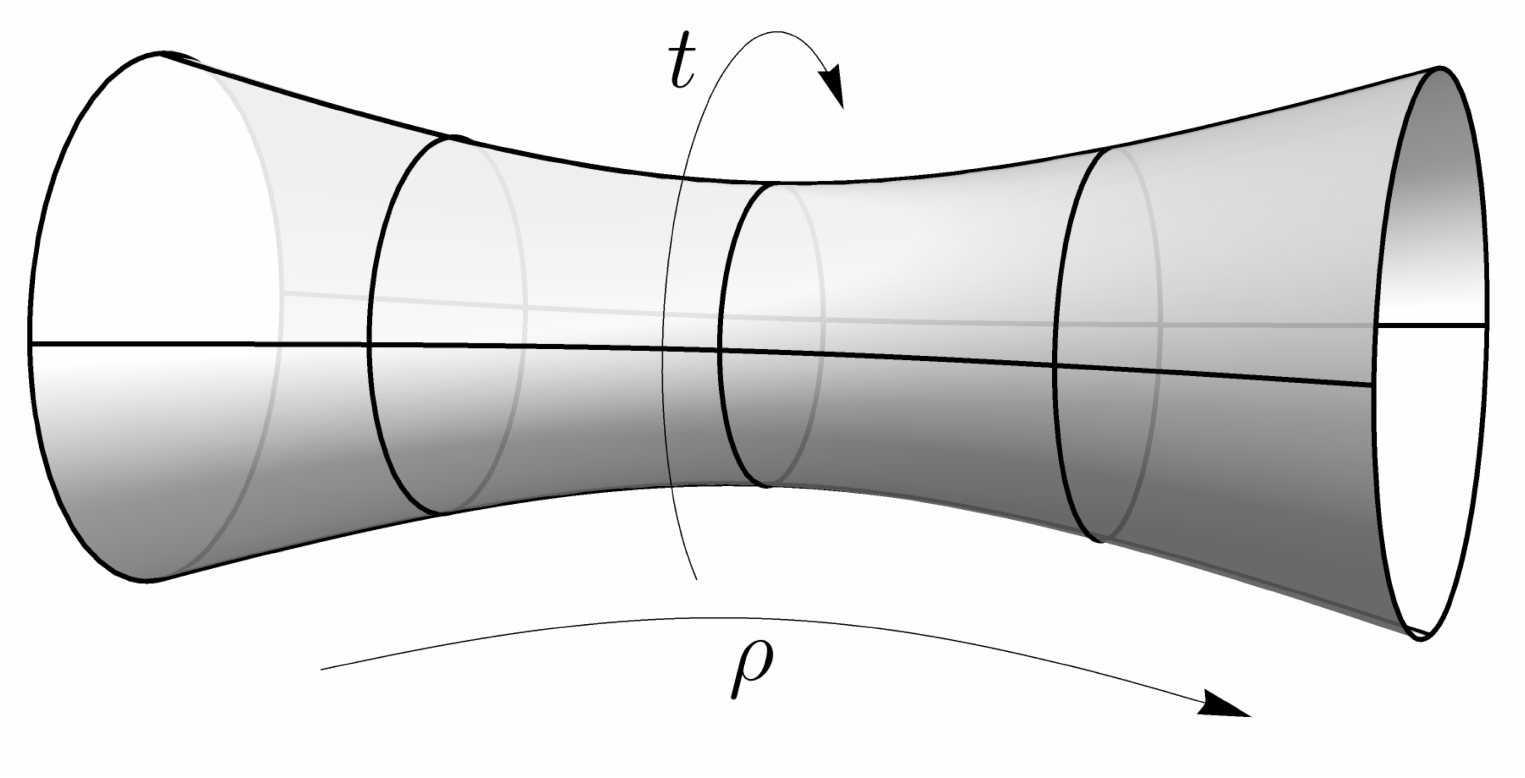}
\caption{A hyperboloid representing $\textrm{AdS}_2$ embedded in flat (1+2)-dimensional space. Time $t$ flows round the neck. The spatial variable $\rho$ goes from $-\pi/2$ to $\pi/2$, with the throat located at $\rho=0$, where the radius is least and given by $L$. The uniformly accelerated trajectories considered in the text correspond to $\rho=\rho_0=const$.}
\label{hyperboloide}
\end{figure}

The metric of  global $\textrm{AdS}_2$ is given by
\begin{equation}
ds^2=\frac{L^2}{\cos^2{\rho}}(-dt^2+d\rho^2),
\end{equation}
where $L$ is the radius of the  hyperboloid, see Fig. \ref{hyperboloide} (we are actually considering here the universal covering of $\textrm{AdS}_2$, with $-\infty<t<\infty$ and $-\pi/2 <\rho<\pi/2$). We restrict ourselves to a conformal scalar field propagating in global $\textrm{AdS}_2$, i.e., a massless minimally coupled field the respecting Klein-Gordon equation
\begin{equation}
\frac{\partial^2 \phi(t,\rho)}{\partial \rho^2}=\frac{\partial^2 \phi(t,\rho)}{\partial t^2},\,\,\,-\pi/2<\rho<\pi/2.
\label{Klein}
\end{equation}
This makes it evident that appropriate boundary conditions are required at $\rho=\pm\pi/2$.

We are interested in uniformly accelerated trajectories in AdS$_2$ given by constant $\rho=\rho_0$ in these coordinates,
\begin{equation}\label{trajectories}
    x(\tau)=\left(\frac{\tau \cos{\rho_0}}{L},\rho_0\right),
\end{equation}
with a subcritical acceleration~\cite{jennings}
\begin{equation}
a^2=a^{\mu}a_{\mu}=\frac{\sin^2{\rho_0}}{L^2}, \,\,\,-\pi/2<\rho_0<\pi/2.
\end{equation}
The trajectory given by $\rho_0=0$ is inertial, while the trajectories represented by $\rho\neq 0$ are accelerated. It is worth noting that both the acceleration and curvature scalar approaches zero when $L\to\infty$, the last of which as
\begin{equation}
    \mathcal{R}=-2L^{-2}.
\end{equation}

\section{Wightman functions}
\label{sec:wightman}

Here we consider the Wightman functions associated with the two classes of boundary conditions described above.

\subsection{\label{sec:sol} Robin boundary conditions}

\begin{figure}[ht]
\centering
\includegraphics[width=0.95\columnwidth]{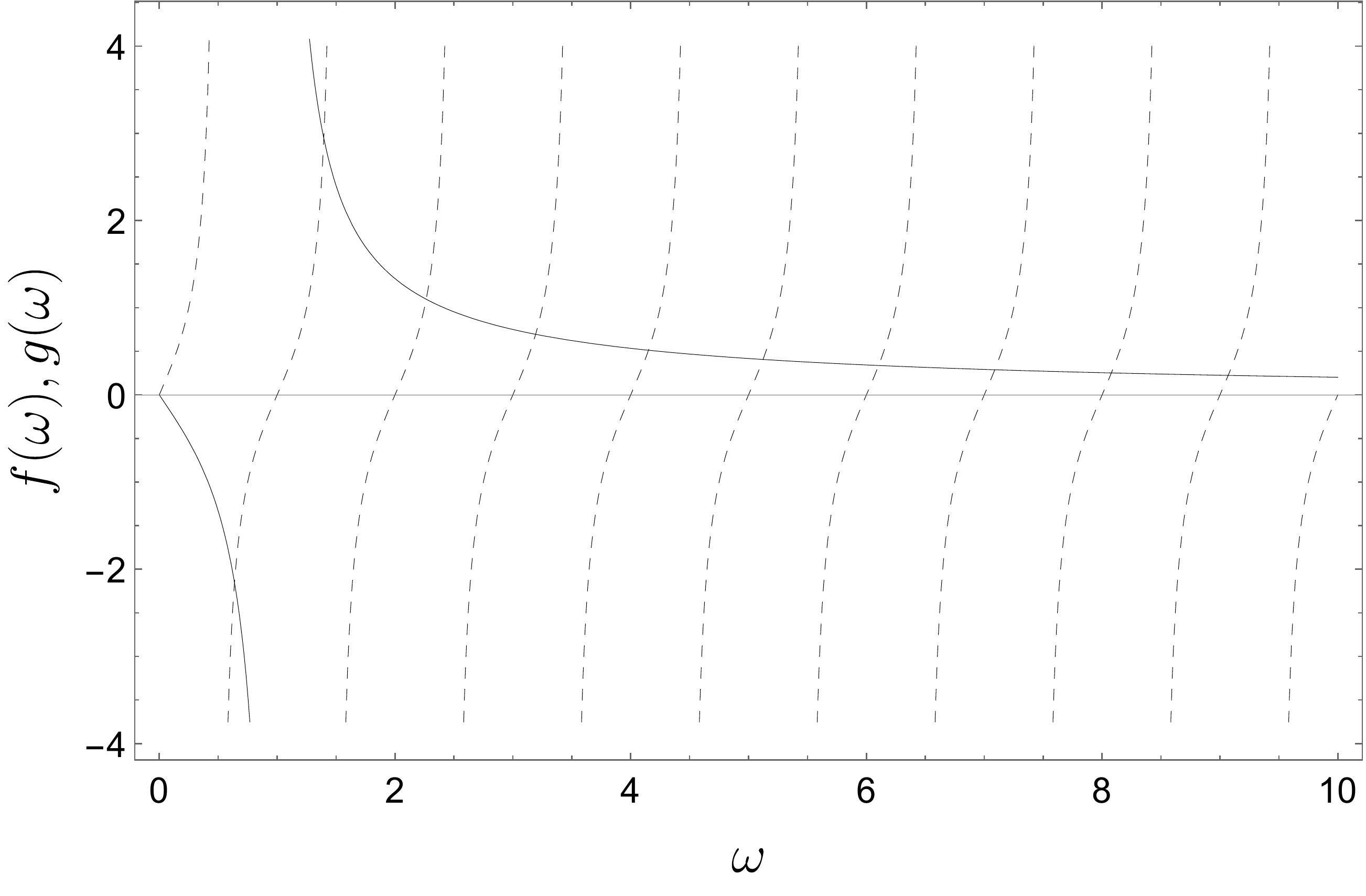}
\caption{The roots of Eq. (\ref{bcrobin}) are given by the intersection of the curve $f(\omega)=\tan{(\pi\omega)}$ (dashed curve) and  $g(\omega)=\frac{2\beta\omega}{\beta^2\omega^2-1}$ (solid curve). The figure illustrates the case of $\beta=1$.}
\label{fig1}
\end{figure}

Robin boundary conditions at both  end points are given by
\begin{equation}
\begin{aligned}
&\phi(t,-\pi/2)-\beta_1\left.\frac{\partial \phi(t,\rho)}{\partial \rho}\right|_{\rho=-\pi/2}=0,\\ &\phi(t,\pi/2)+\beta_2\left.\frac{\partial \phi(t,\rho)}{\partial \rho}\right|_{\rho=\pi/2}=0,
\end{aligned}
\end{equation}
with $\beta_1,\beta_2\in\mathbb{R}$. We choose, for simplicity, $\beta_1=\beta_2=\beta\geq0$, which makes the spatial operator
\begin{equation}
A=-\frac{\partial^2}{\partial \rho^2}
\end{equation}
positive and the system stable~\cite{martino}. The positive frequency solutions in this case are then given by
\begin{equation}\label{modes robin}
u_{\omega_n}^{\beta}=\frac{\sin{\left[\omega_n\left(\rho+\frac{\pi}{2}\right)\right]}+\beta\omega_n\cos{\left[\omega_n\left(\rho+\frac{\pi}{2}\right)\right]}}{\sqrt{\pi \omega_n+2\beta\omega_n+\pi\beta^2\omega_n^3}}e^{-i \omega_n t},
\end{equation}
where $\omega_n$ is the $n$th positive root of the equation
\begin{equation}
\sin{\left(\pi\omega\right)}+\beta\omega\left[2\cos{\left(\pi\omega\right)}-\beta\omega\sin{(\pi\omega)}\right]=0.
\label{bcrobin}
\end{equation}
Figure~\ref{fig1} shows the first roots of Eq. (\ref{bcrobin}) for $\beta=1$. 

We note that the first root, $\omega_0$, tends to zero as $\beta\to\infty$. This illustrates the fact that the Neumann boundary condition has a zero mode~\cite{martinez}, which breaks time translation invariance.  Since this invariance will be essential in what follows, we will not deal with this case. On the other hand, for $n\gg1$, we have $\omega_n\sim n$. Notice that the scale parameter $L$ does not appear in Eq.~(\ref{modes robin}), since we are dealing with a conformal field. In this way, both the Robin parameter $\beta$ and the roots $\omega_n$ are dimensionless.

\subsubsection{The AdS invariant $\beta=0$ (Dirichlet) case}

Since the roots $\omega_n$ satisfy a transcendental equation, a closed form for the Wightman function,
\begin{equation}
W_{\epsilon}(t,\rho;t^{\prime},\rho^{\prime})=\sum_{j}{u_{j}(t-i\epsilon,\rho)u^{\ast}_{j}(t^{\prime},\rho^{\prime})},
\end{equation} 
is only available when $\beta=0$, which corresponds to Dirichlet boundary conditions. We find in this case
\begin{equation}\label{modes dirichlet}
u_n^{\beta=0}=\frac{\sin{\left[n\left(\rho+\frac{\pi}{2}\right]\right)}}{\sqrt{\pi n}}e^{-i n t},\,\,\, n=1,2,3,\dots,
\end{equation}
with the associated Wightman function given by
\begin{equation}\begin{aligned}
&W^{\beta=0}_{\epsilon}(t,\rho;t^{\prime},\rho^{\prime})=\frac{1}{4\pi}\times\\&\ln{\left[\frac{\left(1+e^{-i (\Delta t-\Delta^{+}\rho-i \epsilon)}\right)\left(1+e^{-i (\Delta t+\Delta^{+}\rho-i \epsilon)}\right)}{\left(1-e^{-i (\Delta t+\Delta\rho-i \epsilon)}\right)\left(1-e^{-i (\Delta t-\Delta\rho-i \epsilon)}\right)}\right]},
\end{aligned}
\label{dirichlet bc}
\end{equation}
where $\Delta\rho\equiv \rho-\rho^{\prime}$ and $\Delta^{+}\rho\equiv\rho+\rho^{\prime}$. 

It is worth noting that the (Dirichlet) vacuum is then $\text{AdS}$ invariant~\cite{pitelli1} and that $W^{\beta=0}$ is only a function of the geodesic distance $s$, given by
\begin{equation}
\cosh\left(\frac{s}{L}\right)=1+\frac{s_e^2}{2L^2},
\end{equation}
with $s_e^2=2L^2\left(\cos{\Delta t}-\cos{\Delta\rho}\right)\sec{\rho}\sec{\rho^{\prime}}$~\cite{kent}.  


\subsection{\label{sec:ngh}Pseudoperiodic boundary conditions}

These are given by 
\begin{equation}\begin{aligned}
&\phi(t,\pi/2)=e^{i\theta}\phi(t,-\pi/2),\\ &\left.\frac{\partial \phi(t,\rho)}{\partial \rho}\right|_{\rho=\pi/2}=e^{i\theta}\left.\frac{\partial \phi(t,\rho)}{\partial \rho}\right|_{\rho=-\pi/2}.
\end{aligned}\label{pseudo}
\end{equation}
Within this class of boundary conditions, there is a net flux of information between the end points, as if they were connected like a ring. It can be shown that the self-adjoint extensions in this case are labeled by the phase difference $\theta\in [0,2\pi)$. The cases $\theta=0$ and $\theta=\pi$ correspond to periodic and antiperiodic boundary conditions, respectively.

For $\theta\neq 0,\pi$, the complete set of orthonormal modes satisfying Eqs. (\ref{Klein}) and (\ref{pseudo}) is given by \cite{fn1}
\begin{equation}
\begin{aligned}
 u^{\theta}_n(t,\rho)=&\frac{\exp{\left[i\left(2n+\frac{\theta}{\pi}\right)\left(\rho+\frac{\pi}{2}\right)\right]}}{\sqrt{4n \pi+2\theta}}\times\\&\times \exp{\left[-i\left(2n+\frac{\theta}{\pi}\right)t\right]},\,\,\,\,n=0,1,2,\dots,\\
 v^{\theta}_n(t,\rho)=&\frac{\exp{\left[-i\left(2n-\frac{\theta}{\pi}\right)\left(\rho+\frac{\pi}{2}\right)\right]}}{\sqrt{4n \pi-2\theta}}\times\\&\times\exp{\left[-i\left(2n-\frac{\theta}{\pi}\right)t\right]},\,\,\,\,n=1,2,3,\dots,
\end{aligned}
\label{pseudo general}
\end{equation}
and their complex conjugate. With respect to the timelike Killing field $\partial/\partial t$, $u^{\theta}_{n}$ and $v^{\theta}_n$ are positive frequency modes with energies $2n+\frac{\theta}{\pi}$ and   $2n-\frac{\theta}{\pi}$, respectively. 

For $\theta=0$ and $\theta=\pi$, the spectrum degenerates and the positive frequency modes are given by
\begin{equation}\label{zero and pi}
\begin{aligned}
u^{\theta=0}_n(t,\rho)=&\frac{\exp{\left[2i n \left(\rho+\frac{\pi}{2}\right)\right]}}{\sqrt{2\pi}\sqrt{2|n|}}\times \\&\times\exp{\left[-2i |n| t\right]},\,\,\,n=\pm 1,\pm 2,\pm 3,\dots,\\
u^{\theta=\pi}_n(t,\rho)=&\frac{\exp{\left[(2n+1)i \left(\rho+\frac{\pi}{2}\right)\right]}}{\sqrt{2\pi}\sqrt{|2n+1|}}\times\\&\times \exp{\left[-|(2n+1)|i  t\right]},\,\,\,n=0,\pm 1,\pm 2,\dots.
\end{aligned}
\end{equation}

A closed form for the Wightman functions for $\theta\neq 0, \pi$ can be obtained by the usual representation of the hypergeometric function as a Gauss series~\cite{abramo}. This leads to
\begin{equation}\label{wightman periodic}\small
\begin{aligned}
&W^{\theta}_{\epsilon}(t,\rho;t^{\prime},\rho^{\prime})=\\&\frac{e^{-\frac{i\theta(\Delta t-\Delta \rho-i\epsilon)}{\pi}}}{2\theta}\phantom{}_2F_1\left(1,\frac{\theta}{2\pi};1+\frac{\theta}{2\pi};e^{-2i (\Delta t-\Delta \rho -i \epsilon)}\right)+\\&\frac{e^{-\frac{i(2\pi-\theta)(\Delta t+\Delta \rho-i\epsilon)}{\pi}}}{2(2\pi-\theta)}\phantom{}_2F_1\left(1,1-\frac{\theta}{2\pi};2-\frac{\theta}{2\pi};e^{-2i (\Delta t+\Delta \rho -i \epsilon)}\right).
\end{aligned}
\end{equation}

For the case $\theta=0,\, \pi$, the sum (\ref{zero and pi}) yields [using the expansions of $\ln{(1+z)}$ and $\arctanh{(z)}$ as Taylor series for $|z|<1$]
\begin{equation}\begin{aligned}\small
&W^{\theta=0}_{\epsilon}(t,\rho;t^{\prime},\rho^{\prime})=\\&-\frac{\ln{\left[1-e^{-2i(\Delta t-\Delta\rho-i \epsilon)}\right]}+\ln{\left[1-e^{-2i(\Delta t+\Delta\rho-i \epsilon)}\right]}}{4\pi},\\
&W^{\theta=\pi}_{\epsilon}(t,\rho;t^{\prime},\rho^{\prime})=\\&\frac{\arctanh{\left[e^{-i(\Delta t-\Delta\rho-i \epsilon)}\right]}+\arctanh{\left[e^{-i(\Delta t+\Delta\rho-i \epsilon)}\right]}}{2\pi}.
\end{aligned}
\label{particular cases}
\end{equation}

The manifest time dependence only through $\Delta t=t-t^{\prime}$  shows that the vacua defined by the modes $u_{n}^{\theta}(t,\rho)$ are invariant under time translation given by the Killing field $\partial/\partial t$. Moreover,  even though $\partial/\partial \rho$ is not a Killing field, the Wightman functions are only sensitive to $\Delta \rho=\rho-\rho^{\prime}$, which reflects the fact that the conformal field is propagating effectively on a ring for this choice of boundary conditions.

\section{\label{sec:response}Response of the detector in $\textrm{AdS}_2$}
We proceed to calculate the response function and transition rate associated with the classes of boundary conditions considered above.

\subsection{Dirichlet boundary conditions}

We start with the Dirichlet boundary condition, for which there is no net flux of energy through the conformal boundaries, so that the system spacetime+scalar field is isolated. 

To find the response function, we make use of the Wightman function given by Eq.~(\ref{dirichlet bc}).  Expanding $\ln(1\pm z)$ as a Taylor series around $z=0$, the discrete character of the energy spectrum becomes manifest. It is worth noting how the $\epsilon>0$ prescription is crucial here, as the Taylor series diverges at $z=-1$ [for  $\ln(1+ z)$[ and $z=1$ [for  $\ln(1-z)$]. It follows from Eq.~(\ref{resp}) that
\begin{equation}
        \begin{aligned}
          &\mathcal{F}_{T}^{\beta=0}(\Omega)
          =\lim_{\epsilon\to 0^{+}} \sum_{k=1}^{\infty}2\int_{T_0}^{T}\mathrm{d}u\int_{0}^{u-T_0}\mathrm{d}s \,e^{-\epsilon k}\times\\&\times\frac{\left[1+(-1)^{k+1}\cos{(2k\rho_{0})}\right]\cos{\left[s\left(\Omega +k\frac{ \cos{\rho_{0}}}{L}\right)\right]}}{2 \pi k}\\
            &=\lim_{\epsilon\to 0}\Bigg\{\sum_{k=1}^{\infty}e^{-\epsilon k}\frac{2\left[1+(-1)^{k+1}\cos{(2k\rho_{0})}\right]}{\pi k }\times \\& \times\frac{\sin{\left[\frac{(T-T_0)}{2}\left(\Omega +\frac{k \cos{\rho_{0}}}{L}\right)\right]}^{2}}{ \left(\Omega+\frac{k \cos{\rho_{0}}}{L}\right)^{2}}\Bigg\}.
        \end{aligned} \label{resp dirichlet}
\end{equation}
This series converges uniformly so that we can interchange the infinite sum and the $\epsilon\to 0^{+}$ limit. As a result, the $e^{-\epsilon k}$ term can be set to $1$ and we get
\begin{equation}\begin{aligned}
&\mathcal{F}_{T}^{\beta=0}(\Omega)=\\&\sum_{k=1}^{\infty}\frac{2\left[1+(-1)^{k+1}\cos{(2k\rho_{0})}\right]\sin{\left[\frac{(T-T_0)}{2}\left(\Omega +\frac{k \cos{\rho_{0}}}{L}\right)\right]}^{2}}{ \pi k\left(\Omega+\frac{k \cos{\rho_{0}}}{L}\right)^{2}}.
\end{aligned}
\label{resp dirichlet final}
\end{equation}

We see that $ \mathcal{F}_{T}^{\beta=0}(\Omega)$ is not defined in the limit $T_0\to -\infty$, which is not a surprise as this also happens in Minkowski spacetime. To deal with interactions on an infinite proper time interval and eliminate transient effects, we consider, as usual, the limit $T_0\to-\infty$ in the transition rate  $ \dot{\mathcal{F}}_{T}^{\beta=0}(\Omega)$. It follows from Eq.~(\ref{rate}) that (in the $\epsilon\to 0^+$ limit)
\begin{equation}\label{rate dirchilet1}
   \begin{aligned} &\dot{\mathcal{F}}_{T}^{\beta=0}(\Omega)=\\&\sum_{k=1}^{\infty}\frac{\left[1+(-1)^{k+1}\cos{(2k\rho_{0})}\right]\sin{\left[(T-T_0)\left(\Omega +\frac{k \cos{\rho_{0}}}{L}\right)\right]}}{\pi k  \left(\Omega+\frac{k \cos{\rho_{0}}}{L}\right)}.
\end{aligned}
\end{equation}
Now the limit $T_0\to-\infty$  is well defined and
\begin{equation}\label{rate dirichlet2}
\begin{aligned}
   & \dot{\mathcal{F}}^{\beta=0}(\Omega)=\lim_{T\rightarrow \infty}\dot{\mathcal{F}}_{T}^{\beta=0}(\Omega)=\\&\sum_{k=1}^{\infty}\frac{1+(-1)^{k+1}\cos{(2k\rho_{0})}}{k}\delta\left(\Omega+k\frac{\cos{\rho_{0}}}{L}\right),
\end{aligned}
\end{equation}
where we have used the identity
\begin{equation}
\lim_{a\to\infty}\frac{\sin{a x}}{\pi x}=\delta (x)
\end{equation}
for the Dirac delta function.

Equation~(\ref{rate dirichlet2}) shows that the transition rate is nonzero only when $\Omega<0$. This corresponds to the case when the final energy $E_f$ of the detector is lower than its initial energy $E_i$. In other words, there is deexcitation of the detector, which emits a particle with energy $|\Omega|$ so that the field becomes excited. On the other hand, since the transition rate vanishes for $\Omega>0$, the detector is never excited by the field. We emphasize here that this behavior happens because we are working in the subcritical regime; this would not be the case for supercritical accelerated detectors for which a thermal response is expected~\cite{jennings}. In summary, when $\Omega=-k\cos{\rho_0}/L$, $k=1,2,3,\dots$, the detector has a nontrivial probability of spontaneously emitting a particle with energy $k\cos{\rho_0}/L$.

 For the inertial trajectory, given by $\rho_0=0$, Eq.~(\ref{rate dirichlet2}) reads
\begin{equation}
    \dot{\mathcal{F}}^{\beta=0}(\Omega)=\sum_{k=0}^{\infty}\frac{2}{2k+1}\delta\left(\Omega+\frac{(2k+1)}{L}\right),
\end{equation}
which means that the detector is only allowed to decay by the exchange of an odd energy excitation. A similar calculation shows that, even if the field is initially in an excited state $\left|2k\rangle\right.$, $k=1,2,\dots$, it will never excite the detector. This illustrates how subtle the particle concept is in curved spacetimes. 
By adopting the point of view that a particle is what a detector detects, we see that there is an infinite number of (even energy) excitations of the field that evade detection as particles by this inertial observer.

We suspect that this unexpected behavior reflects the fact that the even energy modes in  Eq.~(\ref{modes dirichlet}) violate parity, which is clearly a symmetry for the inertial detector configuration in $\textrm{AdS}_2$.
Since the Dirichlet vacuum is $\textrm{AdS}$ invariant,  every pointlike  inertial observer  will be unable to exchange even energy excitations with the field.  Given that parity may be violated by spatially extended detectors following a generic timelike geodesic, it would be very interesting to analyze whether an arbitrary inertial observer will be able to ``see'' these missing excitations. We leave this analysis for a future work.

\subsection{The $\beta>0$ case}
For a generic  Robin boundary condition parametrized by $\beta>0$, we can solve Eqs.~(\ref{resp}) and~(\ref{rate}) by considering the Wightman function as a sum of modes. The transition probability and response rate then become
\begin{widetext}
\begin{equation}\label{resp robin}
  \mathcal{F}^{\beta}_{T}(\Omega)=\sum_{n=0}^{\infty}\frac{4\left\{\sin{\left[\omega_{n}\left(\rho_{0}+\frac{\pi}{2}\right)\right]}+\beta \omega_{n} \cos{\left[\omega_{n}\left(\rho_{0}+\frac{\pi}{2}\right)\right]}\right\}^{2}\sin{\left[\frac{(T-T_0)}{2} \left(\Omega+\frac{\omega_{n}\cos{ \rho_{0}}}{L}\right)\right]^{2}}}{(\pi \omega_{n}+2\beta \omega_{n}+\pi \beta^{2}\omega_{n}^{3})\left(\Omega+\frac{\omega_{n} \cos{\rho_{0}}}{L}\right)^{2}}
  \end{equation}
  and 
\begin{equation}
\label{rate robin1}
  \dot{\mathcal{F}}^{\beta}(\Omega)=\sum_{n=0}^{\infty}\frac{2\pi\left\{\sin{\left[\omega_{n}\left(\rho_{0}+\frac{\pi}{2}\right)\right]}+\beta \omega_{n} \cos{\left[\omega_{n}\left(\rho_{0}+\frac{\pi}{2}\right)\right]}\right\}^{2}}{\pi \omega_{n}+2\beta \omega_{n}+\pi \beta^{2}\omega_{n}^{3}}\delta\left(\Omega+\frac{\omega_n\cos{\rho_0}}{L}\right).
\end{equation}
\end{widetext}

We see that, in this case, deexcitation of the detector can only happen when
\begin{equation}
    \Omega=-\frac{\omega_n \cos{\rho_0}}{L},
    \label{energyrobin}
\end{equation}
where $\omega_n$ satisfies Eq.~(\ref{bcrobin}).

\subsection{Pseudoperiodic boundary conditions}

For the pseudoperiodic boundary conditions, we can expand the hypergeometric functions as Taylor series (or, equivalently,  use the sum of modes form for the Wightman function) in Eqs.~(\ref{wightman periodic}) and (\ref{particular cases}). After substituting these expressions in Eqs.~(\ref{resp}) and~(\ref{rate}), we arrive at 
\begin{widetext}
\begin{equation}\label{resp periodic}
   \mathcal{F}_{T}^{\theta}(\Omega)=\sum_{k=0}^{\infty}\frac{2}{\pi}\left\{\frac{\sin^{2}\left[\frac{(T-T_0)}{2}\left(\Omega + \frac{2 k+2-\frac{\theta }{\pi } }{L}\cos{\rho_{0}} \right)\right]}{\left(2k+2-\frac{\theta}{\pi}\right)\left(\Omega + \frac{2 k+2-\frac{\theta }{\pi } }{L}\cos{\rho_{0}} \right)^2}+\frac{\sin^{2}\left[\frac{(T-T_0)}{2}\left(\Omega + \frac{2 k+\frac{\theta }{\pi } }{L}\cos{\rho_{0}} \right)\right]}{\left(2k+\frac{\theta}{\pi}\right)\left(\Omega + \frac{2 k+\frac{\theta }{\pi } }{L}\cos{\rho_{0}} \right)^2}\right\}
\end{equation}
and
\begin{equation}\label{rate periodic}
     \dot{\mathcal{F}}^{\theta}(\Omega)=\sum_{k=0}^{\infty} \left[\frac{1}{2k+2-\frac{\theta}{\pi}}\delta\left(\Omega + \frac{\left(2 k+2-\frac{\theta }{\pi }\right) \cos{\rho_{0}}}{L}\right)+\frac{1}{2k+\frac{\theta}{\pi}}\delta\left(\Omega + \frac{\left(2 k+\frac{\theta }{\pi}\right) \cos{\rho_{0}}}{L}\right)\right].
\end{equation}
\end{widetext}

Once again, an initially excited detector may emit a particle, whose energy spectrum can now be found analytically. We immediately see that this is given by $|\Omega|= \frac{\left(2 k+\frac{\theta }{\pi}\right) \cos{\rho_{0}}}{L}$ and $|\Omega|= \frac{\left(2 k+2-\frac{\theta }{\pi}\right) \cos{\rho_{0}}}{L}$, $k=0,1,2,\dots$, in this case.

It is worth noting that Eqs.~(\ref{rate robin1}) and (\ref{rate periodic}) unequivocally show the nontrivial dependence of the   deexcitation energies with the boundary condition of the field.

\section{\label{sec:limits} The $L\to\infty$ limit}
We proceed to show that the results of the previous section reproduce the well-known transition rate of Minkowski spacetime when $L\to\infty$, irrespective of our choice of boundary conditions. The significance of this result is then analyzed in the next section.

\subsection{\label{sec:dirichlet limit} Dirichlet boundary conditions}

By splitting  Eq.~(\ref{rate dirichlet2}) into two sums, of odd and even terms, respectively, we get
\begin{equation}
\begin{aligned}\dot{\mathcal{F}}^{\beta=0}(\Omega)&=\sum_{k=0}^{\infty}{\frac{2\cos^2{((2k+1)\rho_0)}}{2k+1}}\delta\left(\Omega+(2k+1)\frac{\cos{\rho_0}}{L}\right)\\&+
\sum_{k=1}^{\infty}{\frac{2\sin^2{(2k\rho_0)}}{2k}}\delta\left(\Omega+2k\frac{\cos{\rho_0}}{L}\right).
\end{aligned}
\end{equation}
We define $u\equiv (2k+1)\frac{\cos{\rho_0}}{L}$ and $v\equiv 2k\frac{\cos{\rho_0}}{L}$ in the first and second lines of the above equation, respectively. This yields
\begin{equation}
\label{eq:delta_u}
\begin{aligned}
&1=\Delta k=\Delta u\frac{L}{2\cos{\rho_0}},\\
&1=\Delta k=\Delta v\frac{L}{2\cos{\rho_0}},
\end{aligned}
\end{equation}
so that
\begin{equation}
\begin{aligned}\dot{\mathcal{F}}^{\beta=0}(\Omega)=&\sum_{u=\cos{\rho_0}/L}^{\infty}{\frac{L\Delta u}{\cos{\rho_0}}\frac{\cos^2\left(\frac{uL}{\cos{\rho_0}}\right)}{uL/\cos{\rho_0}}}\delta\left(\Omega+u\right)+\\&
\sum_{v=\cos{\rho_0}/(2L)}^{\infty}{\frac{L\Delta v}{\cos{\rho_0}}}\frac{\sin^2\left(\frac{vL}{\cos{\rho_0}}\right)}{vL/\cos{\rho_0}}\delta\left(\Omega+v\right).
\end{aligned}
\end{equation}
In the $L\to\infty$ limit, each sum with steps $\Delta u$ and $\Delta v$ turns into  an integral in $u$ and $v$,  respectively.  Combining the integrals, we get,  as a result, 
\begin{equation}
\dot{\mathcal{F}}^{\beta=0}(\Omega)=\int_{0}^{\infty}{\frac{\delta(\Omega+\omega)}{\omega}d\omega}=-\frac{1}{\Omega}\Theta(-\Omega).
\end{equation}
This is precisely the transition rate for an inertial detector in Minkowski spacetime (\ref{ratefinal}).

\subsection{\label{sec:robin limit} Robin boundary conditions}

Unfortunately, the calculation above is not easily generalizable for a Robin boundary condition parametrized by $\beta>0$. However, we show next by a semianalytical procedure that the same result as before holds here.

\begin{figure}[ht]
\centering
\includegraphics[width=0.9\columnwidth]{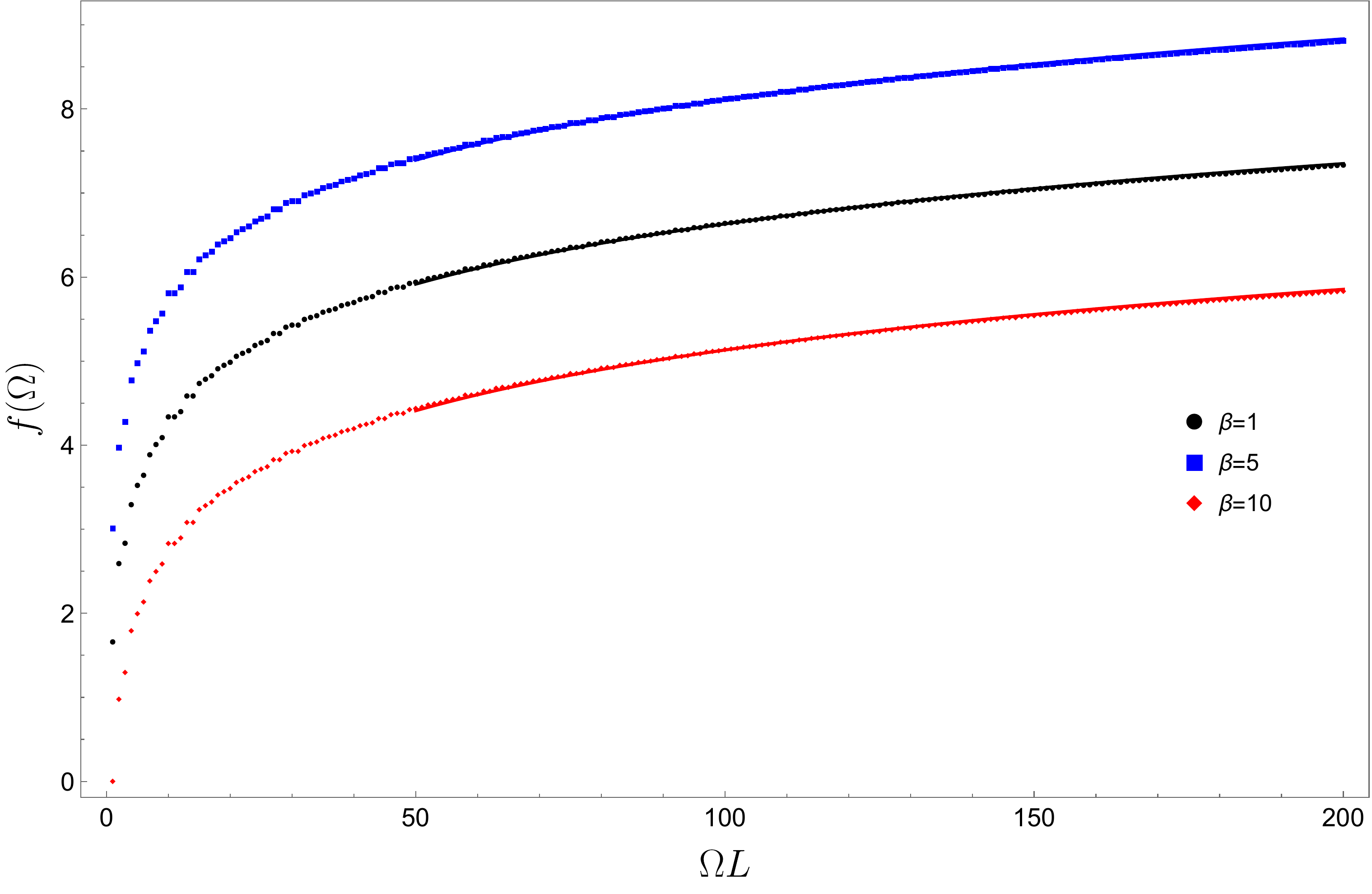}
\caption{Comparison between the values of the step function $f(\Omega)$ given by Eq.~(\ref{robin integrado}) and the continuous function $\tilde{f}(\Omega)$ (solid curves) for $\rho_0=\pi/4$ and $\beta=1,5,10$. The constant term in Eq.~(\ref{fit}) was numerically found by fitting $\tilde{f}(\Omega)$ to $f(\Omega)$.}
\label{fig:robin integrado}
\end{figure}

By integrating $\dot{\mathcal{F}}^{\beta}(\Omega)$ given by Eq.~(\ref{resp robin}) from  $\Omega<0$ to $\Omega=0$, we obtain the step function $f(\Omega)$ given by
\begin{equation}\label{robin integrado}
\begin{aligned}
&f(\Omega)=\int_{\Omega}^{0}\dot{\mathcal{F}}^\beta(\tilde{\Omega})d\tilde{\Omega}\\&=\sum_{\omega=\omega_0}^{\lfloor \frac{-\Omega L}{\cos{\rho_0}} \rfloor}\frac{2\pi\left\{\sin{\left[\omega_{n}\left(\rho_{0}+\frac{\pi}{2}\right)\right]}+\beta \omega_{n} \cos{\left[\omega_{n}\left(\rho_{0}+\frac{\pi}{2}\right)\right]}\right\}^{2}}{\pi \omega_{n}+2\beta \omega_{n}+\pi \beta^{2}\omega_{n}^{3}},
\end{aligned}
\end{equation}
where $\lfloor x \rfloor$ is the floor function and $\omega_n$ can be obtained numerically  from Eq.~(\ref{bcrobin}). When $\Omega  L \gg1$, the heights and widths of each step tend to zero so that  $f(\Omega)$ can be approximated by a continuous function $\tilde{f}(\Omega)$. 
We numerically show in Fig.~\ref{fig:robin integrado} that this continuous function is actually given by
\begin{equation}
f(\Omega)\sim \tilde{f}(\Omega)=\ln{\left(\frac{-L \Omega}{\cos{\rho_0}}\right)}+\textrm{constant},\quad \Omega L\gg1,
\label{fit}
\end{equation}
where the constant in the above equation depends on $\beta$,  $L$, and $\rho_0$,  but not on $\Omega$. 

Moreover, $f(\Omega)$ is clearly zero for $\Omega>0$ from the previous section, since there is no spontaneous excitation. 
By taking the derivative of $\tilde{f}(\Omega)$,  we recover the transition rate for an inertial detector in 1+1 Minkowski spacetime,
\begin{equation}
    \dot{\mathcal{F}}^{\beta}(\Omega)=-\frac{1}{\Omega}\Theta(-\Omega).
\end{equation}

\subsection{\label{sec:periodic limit} Pseudoperiodic boundary conditions}

Proceeding by analogy with the Dirichlet case, we define $u \equiv \frac{(2k+2-\frac{\theta}{\pi})}{L}\cos{\rho_0}$ and $v\equiv \frac{(2 k+\frac{\theta }{\pi})}{L}\cos{\rho_0}$. This yields again Eqs.~(\ref{eq:delta_u}) which, when substituted into Eq.~(\ref{rate periodic}), leads to
\begin{equation}
\begin{aligned}\dot{\mathcal{F}}^{\theta}(\Omega)=&\sum_{u=(2-\frac{\theta}{\pi})\frac{\cos{\rho_0}}{L}}^{\infty}\frac{L\Delta u}{2\cos{\rho_0}}\frac{\cos{\rho_0}}{uL}\delta\left(\Omega+u\right)+\\&
\sum_{v=\frac{\theta}{\pi}\frac{\cos{\rho_0}}{L}}^{\infty}\frac{L\Delta v}{2\cos{\rho_0}}\frac{\cos{\rho_0}}{vL}\delta\left(\Omega+v\right).
\end{aligned}
\end{equation}
As a result, the limit $L \to \infty$ once again yields
\begin{equation}
    \dot{\mathcal{F}}^{\theta}(\Omega)=\int_{0}^{\infty}{\frac{\delta(\Omega+\omega)}{\omega}d\omega}=-\frac{1}{\Omega}\Theta(-\Omega).
\end{equation}

\section{\label{sec:regularization} $\text{AdS}$ as a natural infrared regulator}

It is well known that the calculation of the Wightman function for the massless scalar field in 1+1 Minkowski spacetime suffers from an inherent infrared ambiguity. As we briefly reviewed in Sec.~\ref{sec:minkowski}, the response rate for the Unruh-DeWitt detector in this case can only be calculated by means of an arbitrary infrared frequency cutoff $m_0$. On the other hand, we have seen in the previous section that the conformal scalar field in AdS$_2$ is free from this ambiguity and recovers the Minkowski result in the limit of $L\to\infty$. It must be the case, then, that the energy scale $1/L$ is effectively playing the role of $m_0$ as an infrared regulator. We show next that this is indeed the case and that, apart from some numerical factors depending on the specific set of boundary conditions for AdS$_2$, $1/L$ acts exactly the same way as $m_0$ to that end.

In order to simplify our analysis, we henceforth take $T_0=-T$ (this choice does not affect our results since we are only  considering translationally invariant configurations).

\subsection{\label{limit dir} Dirichlet boundary condition}
Restoring the units in $\Delta t$ and $\epsilon$ by writing $\Delta t=\Delta \tau\cos{\rho_0}/L$ and $\epsilon=\bar{\epsilon} \cos{\rho_0/L}$ in Eq.~(\ref{dirichlet bc}), and considering  the $\Delta \tau/L\to 0$  limit, we have the following expansions for the arguments of the logarithms in Eq.~(\ref{dirichlet bc}):
\begin{equation}
\begin{aligned}
    &1\pm e^{-i \left(\Delta t-\Delta^{+}\rho-i\epsilon\right)}\sim\\
    &\sim 1\pm e^{-2i \rho_0}\left(1+\frac{i\Delta \tau \cos{\rho_0}}{L}-\frac{\bar{\epsilon} \cos{\rho_0}}{L}\right),\\
     &1\pm e^{-i \left(\Delta t-\Delta\rho-i\epsilon\right)}\sim\\
    &\sim 1\pm \left(1+\frac{i\Delta \tau \cos{\rho_0}}{L}-\frac{\bar{\epsilon} \cos{\rho_0}}{L}\right).
\end{aligned}
\end{equation}
Since the Wightman function has the form
\begin{equation}
\begin{aligned}
&W^{\beta=0}_{\epsilon}=-\frac{1}{2\pi}\times\\ &\times\ln{\left(\frac{\left[1 +e^{-i \left(\Delta t-\Delta^{+}\rho-i\epsilon\right)}\right]\left[1 -e^{-i \left(\Delta t+\Delta^{+}\rho-i\epsilon\right)}\right]}{\left[1+ e^{-i \left(\Delta t-\Delta\rho-i\epsilon\right)}\right]\left[1-e^{-i \left(\Delta t+\Delta\rho-i\epsilon\right)}\right]}\right)^{-\frac{1}{2}}},
\end{aligned}
\end{equation}
we have
\begin{equation}
   W^{\beta=0}_{\epsilon}\sim -\frac{\ln{\left[\frac{1}{2L}\left(\bar{\epsilon}+i\Delta \tau\right)\right]}}{2\pi}.
\end{equation}
Comparing the above equation with Eq.~(\ref{greenmink1}), we conclude that
\begin{equation}\label{regulator dir}
    m_0\sim\frac{1}{2L},
\end{equation}
i.e., $1/2L$ plays, in this context, the exact same role as the infrared frequency cutoff $m_0$. Notice that this regulator is independent of the value of $\rho_0$.

To further check this result, we numerically consider the case of finite time in Fig.~\ref{figtripla}(a), where we plot $\mathcal{F}^{\beta=0}_T(\Omega)$ given by Eq.~(\ref{resp dirichlet final}) and $\mathcal{F}^{M}_T(\Omega)$ given by Eq.~(\ref{transitionfinal}), with $2T=2\pi$ and $L=200\gg2T$.

\subsection{Robin boundary condition}

\begin{figure}[ht]
\centering
\includegraphics[width=0.9\columnwidth]{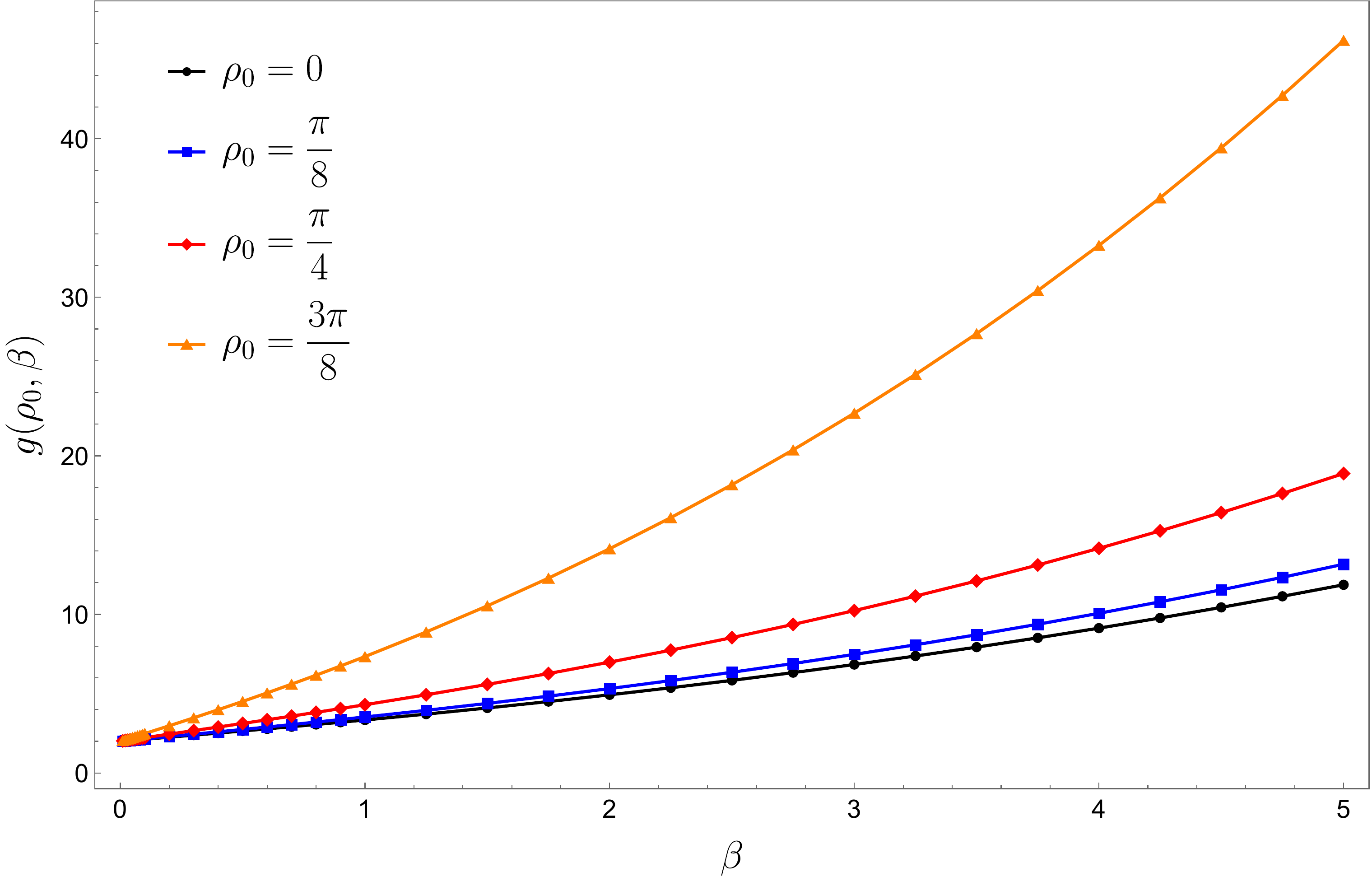}
\caption{The plot shows $g(\rho_0,\beta)$ as a function of $\beta$ for $\rho_0=0,\pi/8,\pi/4$, and $3\pi/8$ [recall that $m_0(\rho_0,\beta)= \frac{1}{g(\rho_0,\beta)L}$]. }
\label{figg}
\end{figure}

For generic Robin boundary conditions, we do not have the Wightman function in closed form. This makes the procedure above impracticable. However, we can still find the relationship between $m_0$ and $1/L$ by resorting to numerics. For each value of $\rho_0$ and $\beta$, we define $m_0(\rho_0,\beta)= \frac{1}{g(\rho_0,\beta)L}$. We then find $g(\rho_0,\beta)$ by fitting the truncated sum in Eq.~(\ref{resp robin}) with the Minkowski response function given by Eq.~(\ref{transitionfinal}). In Fig.~\ref{figg}, we plotted $g(\rho_0,\beta)$ as a function of $\beta$ for several values of $\rho_0$. Notice that in all cases $g(\rho_0,\beta)\to 2$ as $\beta\to 0$, as expected by Eq.~(\ref{regulator dir}). We also note that $g(\rho_0,\beta)$ grows faster when we are closer to the conformal boundary $\rho_0=\pi/2$.

As an example, we choose $\beta=1$ and $\rho_0=\pi/4$. The sum (\ref{resp robin}) produces the continuous curve in Fig.~\ref{figtripla}(b). We see that the Minkowski response function (dashed curve) given by Eq.~(\ref{transitionfinal}) adjusts very well to these points when $m_0$ is given by
\begin{equation}\label{regulator robin}
    m_0\sim\frac{1}{4.3 L}.
\end{equation}

\subsection{Pseudoperiodic boundary condition}
\begin{figure}[ht]
\centering
\includegraphics[width=0.65\columnwidth]{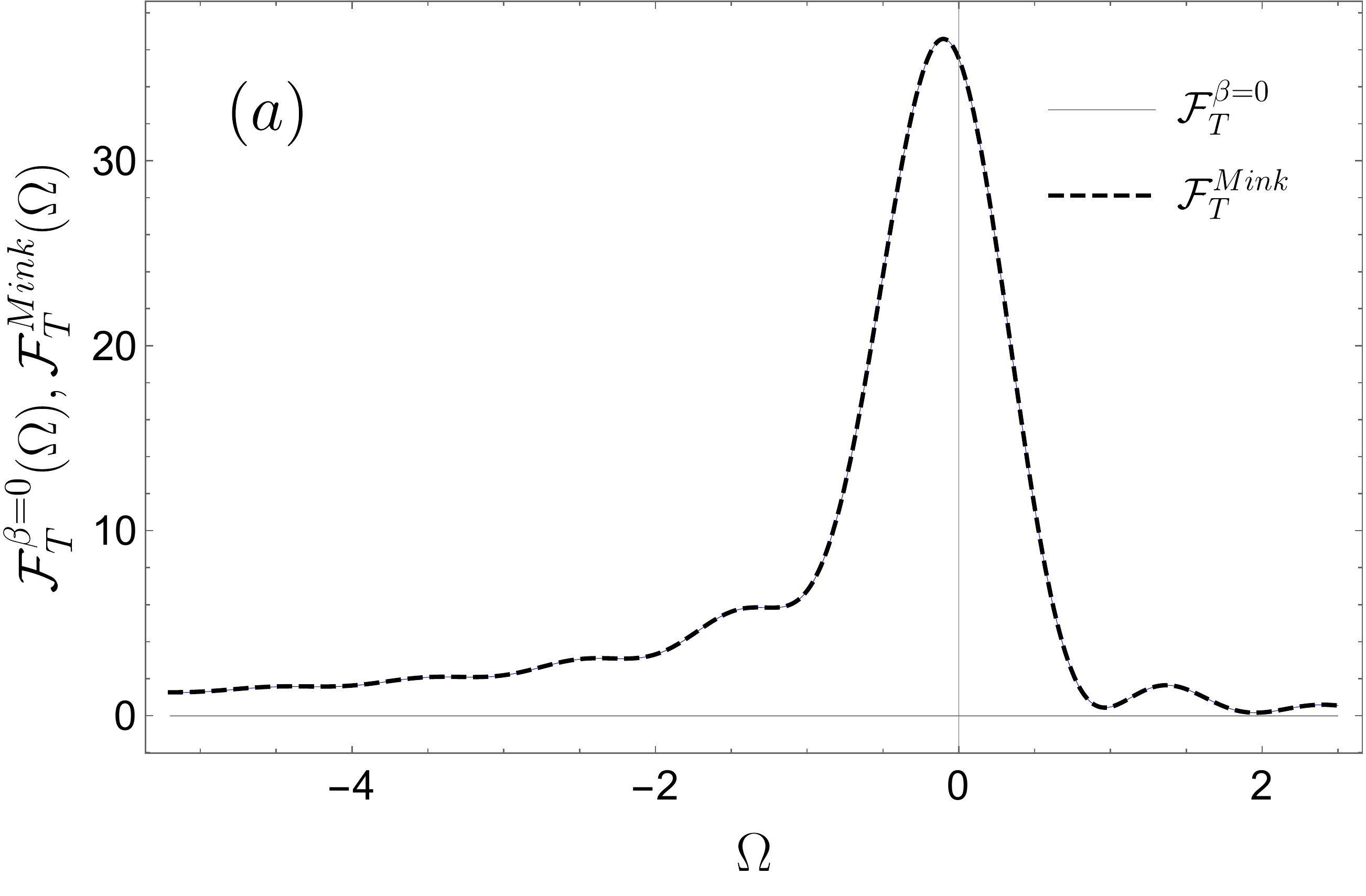}
\includegraphics[width=0.65\columnwidth]{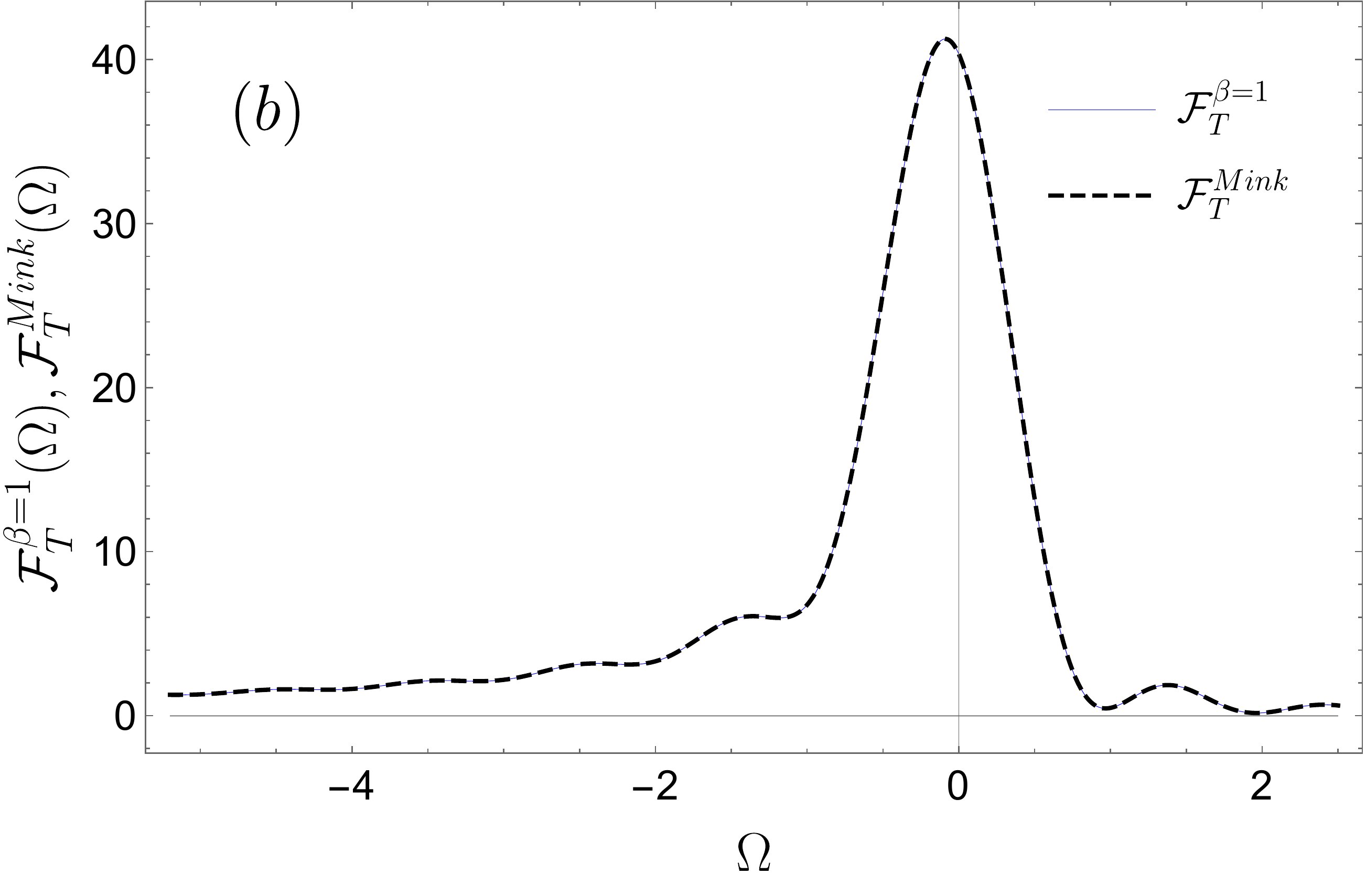}
\includegraphics[width=0.65\columnwidth]{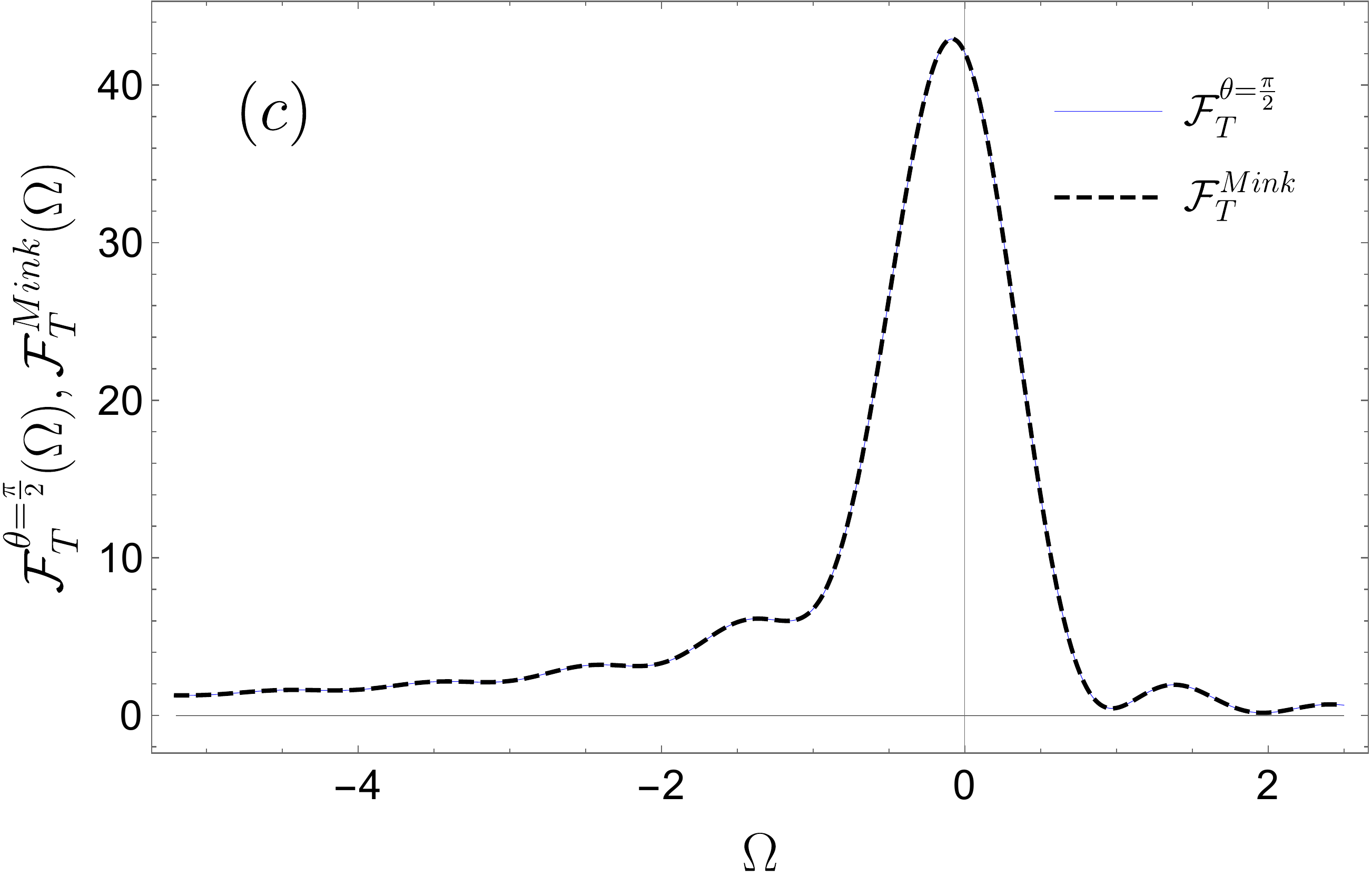}
\caption{Comparison between the response function in $\textrm{AdS}_2$ (solid curves) in the regime of $L\gg2T$ and the response function in 1+1 Minkowski space (dashed curves), which depends on the IR cutoff $m_0$ (see text). In all cases, we considered $L=200$ and $2T=2\pi$ so that $L\gg2T$. (a) The Dirichlet case ($\beta=0$) along with the Minkowski result with $m_0$  given by Eq.~(\ref{regulator dir}). (b) The case of a generic Robin boundary condition by considering $\beta=1$ and $m_0$ given by Eq.~(\ref{regulator robin}). The AdS curve was obtained by computing a truncated sum  in Eq.~(\ref{resp robin}). (c) The case of pseudoperiodic boundary conditions by considering $\theta=\pi/2$ and $m_0$ given by Eq.~(\ref{regulator periodic}).}
\label{figtripla}
\end{figure}

This case may be analyzed by writing $\Delta t=\Delta \tau\cos{\rho_0}/L$ and $\epsilon=\bar{\epsilon} \cos{\rho_0/L}$ in Eq.~(\ref{wightman periodic}). We can then make use of the formula~\cite{abramo}
\begin{equation}
\begin{aligned}
    &\phantom{}_2F_1(a,b;a+b,z)=\\
    &-\frac{\Gamma(a+b)\left[\ln(1-z)+\psi(a)+\psi(b)+2\gamma\right]}{\Gamma(a)\Gamma(b)},
\end{aligned}
\end{equation}
for $|\textrm{arg}(1-z)|<\pi$, $| 1-z|< 1$, where $\psi(x)$ is the polygamma function and $\gamma$ is the Euler-Mascheroni constant, and then consider the limit $\Delta \tau/L\to 0$. This leads to
\begin{equation}
    W^{\theta}_{\epsilon}\sim-\frac{ 2\ln{\left[\frac{2(\bar\epsilon+i\Delta \tau)\cos{\rho_0}}{L}\right]}+\psi\left(\frac{\theta}{2\pi}\right)+H_{-\frac{\theta}{2\pi}}+\gamma}{4\pi},
\end{equation}
where $H_n$ is the harmonic number. Comparing this result to Eq.~(\ref{dirichlet bc}) yields
\begin{equation}
    m_0\sim\frac{2\, \exp{\left[\frac{1}{2}\left(\psi\left(\frac{\theta}{2\pi}\right)+H_{-\frac{\theta}{2\pi}}+\gamma\right)\right]}\cos{\rho_0}}{L}.
    \label{regulator periodic}
\end{equation}
We see that, in this case, the regulator has a nontrivial dependence on $\rho_0$. Figure~\ref{figtripla}(c) shows the response function given by Eq.~(\ref{resp periodic}) along with the response function for the Minkowski spacetime with regulator given by Eq.~(\ref{regulator periodic}). 

\section{Conclusions}
\label{sec:conlusion}

We studied the response of the Unruh-DeWitt detector coupled to a conformal  scalar field in $\textrm{AdS}_2$ spacetime. In particular, we calculated the transition probability and the transition rate for two classes of boundary conditions at the conformal infinities of $\textrm{AdS}_2$, namely Robin and pseudoperiodic boundary conditions. In both cases, the spatial part of the Klein-Gordon equation turns out to be positive and self-adjoint so that the associated field quantization is well defined.

We showed that the transition rate for a detector switched on in the infinite past is given by a sequence of delta functions. These delta functions are supported on a discrete set that depends on the quantum number that characterizes the field modes, on the detector's acceleration, on the $\textrm{AdS}$ energy scale $L$, and on the boundary conditions. We showed that when the energy scale $1/L$ approaches zero the transition rate for an inertial detector in 1+1 Minkowski spacetime is recovered, irrespective of our choice of boundary condition.

A similar conclusion was drawn in Ref.~\cite{henderson} for  $\textrm{AdS}_3$   with  a massless field satisfying Dirichlet, Neumann, and transparent boundary conditions. However, we showed that in $\textrm{AdS}_2$ the Wightman function resembles its Minkowski counterpart, with the mass scale playing the role of an IR regulator. This gives us  a deeper understanding on how and why these limits work. 

Finally, the same idea may be applied to supercritical accelerated detectors, where one should recover the usual thermal spectrum for the Unruh effect. We leave this analysis for a future work.

\acknowledgments
The authors acknowledge insightful discussions with L. de Souza Campos and interesting comments of an anonymous referee. B. S. F.  acknowledges support from the Coordenação de Aperfeiçoamento de Pessoal de Nível Superior (CAPES, Brazil), Grant No. 88887.371836/2019-00. R. A. M. was partially supported by Conselho Nacional de Desenvolvimento Científico e Tecnológico (CNPq, Brazil) under Grant No. 310403/2019-7.


\begin{thebibliography}{99}
\bibitem{fn1}
The solutions given by Eqs.~(\ref{pseudo general}) and (\ref{zero and pi}) were obtained in detail in Ref.~\cite{bonneau}, where the problem of a particle in a box was carefully considered (recall that the conformal field in $\text{AdS}_2$ behaves like a free particle in a box --- see Sec.~\ref{sec:bc}).
\end{thebibliography}
\end{document}